\documentclass[
twocolumn,
12pt,
reprint,
showpacs,preprintnumbers,
showkeys,
amsmath,amssymb,
aps,
prb,
]{revtex4-1}

\usepackage{graphicx}
\usepackage{dcolumn}
\usepackage{bm}
\usepackage{hyperref}
\usepackage{nicefrac}
\usepackage{color}
\usepackage{braket}
\usepackage{multirow}
\usepackage{verbatim}
\usepackage{bbold}
\usepackage[caption=false,subrefformat=parens,labelformat=parens]{subfig}

\newcommand{\LIA}{\lambda_{I}^{A}}
\newcommand{\LIB}{\lambda_{I}^{B}}
\newcommand{\LI}{\lambda_{I}}
\newcommand{\LR}{\lambda_{R}}

\newcommand{\K}{K}
\newcommand{\Kp}{K^\prime}
\newcommand{\B}{\mathcal{B}}
\renewcommand\vec[1]{\textit{\textbf{#1}}}

\makeatletter
\def\maketag@@@#1{\hbox{\m@th\normalfont\normalsize#1}}
\makeatother

\begin{document}

\title{Landau levels in spin-orbit coupling proximitized graphene: bulk states}

\author{\surname{Tobias} Frank}
\email[Emails to: ]{tobias.frank@physik.uni-regensburg.de}
\author{\surname{Jaroslav} Fabian}

\affiliation{%
 Institute for Theoretical Physics, University of Regensburg,\\
 93040 Regensburg, Germany
 }%

\date{\today}

\begin{abstract}
We study the magnetic-field dependence of Landau levels in graphene proximitized by large spin-orbit coupling materials, such as transition-metal dichalcogenides or topological insulators. In addition to the Rashba coupling, two types of intrinsic spin-orbit interactions,  uniform (Kane-Mele type) and staggered (valley Zeeman type), are included, to resolve their interplay with magnetic orbital effects.  Employing a continuum model approach, we derive analytic expressions for low-energy Landau levels, which can be used to extract local orbital and spin-orbit coupling parameters from scanning probe spectroscopy experiments. We compare different parameter regimes to identify fingerprints of relative and absolute magnitudes of intrinsic spin-orbit coupling in the spectra. The inverted band structure of graphene proximitized by WSe$_2$ leads to an interesting crossing of Landau states across the bulk gap at a crossover field, providing insights into the size of Rashba spin-orbit coupling. Landau level spectroscopy can help to resolve the type and signs of the intrinsic spin-orbit coupling by analyzing the symmetry in energy and number of crossings in the Landau fan chart. Finally, our results suggest that the strong response to the magnetic field of Dirac electrons in proximitized graphene can be associated with extremely large self-rotating magnetic moments.
\end{abstract}

\maketitle

\section{Introduction}
The modification of band structures of two-dimensional materials by proximity effect offers a way to design systems with novel properties. In particular for spintronics in graphene~\cite{Han2014} this turns out to be a very promising way to enhance spin-orbit coupling~(SOC) or introducing exchange interaction by proximity with other materials~\cite{Frank2016, Zollner2016a, Gmitra2016}. In this way we can harvest the best from both materials, on the one hand the excellent electronic transport properties of graphene, paired with, e.g., the special optical properties of transition metal dichalcogenides~(TMDs)~\cite{Gmitra2015}. TMDs are two-dimensional van der Waals insulators and are a perfect fit to graphene, because the Dirac cone is situated inside the TMD band gap and leaves the Dirac cone mainly intact. Due to the strong native SOC in TMDs on the order of hundreds of meV, via hybridization processes, the SOC can transfer to graphene.

By now these spin-orbit coupling proximity effects are no longer abstract theoretical concepts, as they have been demonstrated in many experiments~\cite{Garcia2018}. Due to the breaking of spatial symmetries by the substrate, the fundamental properties and spin degeneracy of the graphene Hamiltonian can be altered and a gap be introduced in the originally massless dispersion. Interpretation of experimental data~\cite{Cummings2017a, Ghiasi2017, Benitez2018} hints towards predominant induction of sublattice resolved intrinsic spin-orbit coupling, so-called valley Zeeman intrinsic spin-orbit coupling, where intrinsic spin-orbit coupling is of same magnitude but of different signs in the two sublattices. Other types of effects, such as the breaking of the sublattice symmetry via a staggered potential or enhancement of Rashba spin-orbit coupling are also possible~\cite{Kochan2017}.

Determination of model parameters for symmetry-broken graphene by means of density functional theory~(DFT) calculations was only carried out for commensurate lattice arrangements, but in real systems, lattice mismatch effects could play an important role. For example, the staggered sublattice potential may effectively average out, but could be sizeable locally due to the formation of moir\'{e} patterns~\cite{Li2019, David2019}. It is thus important to devise more direct methods than weak antilocalization and spin precession measurements to explore the \textit{local} electronic processes and energy scales. This could be, for example, achieved by scanning tunneling spectroscopy supplemented with a transverse magnetic field~\cite{Li2007, Yin2015}. It was demonstrated earlier~\cite{Bindel2016} that this method can be used to extract local Rashba parameters in two-dimensional electron gases by comparison with Landau level models. In this spirit, we focus on single-particle Landau level spectra (with approximate $\sqrt{\B}$ magnetic field behavior) rather than on transport signatures (with approximate linear-$\B$ behavior)~\cite{Cysne2018}.

The physics of symmetry-broken honeycomb lattices has been extensively studied theoretically~\cite{Rashba2009, DeMartino2011, Tabert2013, Tsaran2014, Cysne2018}. However, the interplay between orbital magnetic fields and proximity SOC is a relatively recent subject; see, for example, Ref. \onlinecite{Cysne2018} for monolayer and Ref.~\onlinecite{Khoo2018} for bilayer graphene. Due to graphene's linear energy dispersion, the Landau levels of Dirac electrons follow a different magnetic field ($\B$) dependence than in two-dimensional electron gases~\cite{Novoselov2005, Zheng2002, Brey2006}. The graphene Landau levels are given by $\varepsilon_n(\B) = \mathrm{sgn}(n) \sqrt{2v_F^2e\hbar|n|\B}$ ($n = 0, \pm 1, \pm 2, \dots$), where $v_F$ is the Fermi velocity. In the absence of Zeeman coupling, the levels are fourfold degenerate due to the spin and valley degrees of freedom. The Landau level with orbital index $n=0$ is special, being located solely in one sublattice, depending on the valley degree of freedom~\cite{Tabert2015}. When symmetries in a hexagonal lattice become broken as in, e.g., buckled silicene, the fourfold Landau level degeneracy is lifted~\cite{Tabert2013} and gaps can be introduced. While spin-orbit coupling proximitized graphene has many similarities to silicene, in proximitized graphene intrinsic SOC is sublattice resolved and Rashba SOC is momentum independent~\cite{Rashba2009, Tsaran2014, Gmitra2016}. The interplay of uniform intrinsic and Rashba SOC with orbital magnetic fields was studied in Refs.~\onlinecite{DeMartino2011, Rashba2009}. Rashba~\cite{Rashba2009} finds analytic solutions for the Landau states which have in-plane spin-orbit texture and whose hallmark is a twofold degenerate zero energy Landau level. De Martino et al. confirm this result and provide additional analytic results for uniform intrinsic SOC and study magnetic field edge state physics, where they observe phase transitions between different spin-filtered edge regimes. In Ref.~\onlinecite{Island2019} it was shown that using Landau level spectroscopy the magnitude of induced intrinsic SOC can be extracted in proximitized bilayer graphene systems employing capacitance measurements.

In a recent work~\cite{Cysne2018} properties of proximitized graphene under the influence of orbital magnetic fields was studied, where emphasis is put on calculations of transport fan diagrams and Hall conductivities\footnote{A graphene proximity model very similar to ours is employed, where we use another convention~\cite{Kochan2017}, $\K\rightarrow \Kp$, $\LI^{A/B} \rightarrow -\LI^{A/B}$}. This work finds very rich fan diagrams with characteristic level splittings, allowing one to infer the relative sizes of Rashba and intrinsic SOC from scaling observations. The focus is on the behavior of higher-lying Landau levels. 

Our work addresses specifically low-energy and low-magnetic field states which are accessible to scanning tunneling experiments. Additionally, we provide analytic expressions for all four low-energy Landau levels in high magnetic fields. Finally, we also consider the interplay of the valley Zeeman spin-orbit interaction with a staggered potential, which is important to distinguish non-inverted and inverted spectral regimes.

We use a Hamiltonian for the bulk Landau level spectrum of SOC proximitized graphene, including Zeeman interaction, with parameters from DFT predictions~\cite{Gmitra2016}. We show that spectral features of this Hamiltonian may be used to extract important model parameters from experiment by applying magnetic fields up to 15~Tesla. From the model Hamiltonian we extract high-$\B$ field formulas for the energetic behavior of low-energy Landau levels, comparing them with numerical data. Excellent correspondence is found even for smaller values of magnetic fields. These formulas could provide a direct way to measure local intrinsic sublattice-resolved spin-orbit coupling as well as the local sublattice potential by means of Landau level scanning tunneling spectroscopy ~\cite{Li2007, Yin2015, Bindel2016}. Important, this would also provide a direct way to determine whether the intrinsic spin-orbit coupling in proximitized heterostructures is of uniform~\cite{Wakamura2019} or valley Zeeman type~\cite{Wang2015, Yang2016}, differentiating between topologically trivial and nontrivial regimes. These regimes can also be identified by the number of crossings of the low-energy Landau levels in the fan diagram as well as from their electron-hole symmetry.

In the quantum valley spin Hall~(QVSH) regime~\cite{Frank2018}, in which the graphene spin-orbit coupling gap is inverted, we find unique signatures of the gap inversion in the Landau level spectrum. In general, strong orbital magnetic response at low $\B$-fields is accompanied by a crossing of specific Landau levels from the electron to the hole sectors and vice versa. The crossing of states happens at a critical magnetic field, lying in the mT range, for which we provide a formula. The formula offers a method to extract the Rashba spin-orbit coupling strength. The strong magnetic field dependence of those crossing states can be interpreted in terms of magnetic moments, which as we show, are on the order of 1000~$\mu_B$ for realistic parameters. This finding is further corroborated by explicit calculation of the self-rotating magnetic moments for proximitized graphene structures.

The paper is organized as follows. In Sec.~\ref{sec:continuum} we introduce the proximity Hamiltonian and show its form in a magnetic field. Low-energy levels for large magnetic fields are derived and compared to realistic proximity scenarios. Orbital magnetic moments induced by the band structure of proximitized graphene are discussed in Sec.~\ref{sec:magnetic_moments} before concluding with Sec.~\ref{sec:conclusion}.

\section{\label{sec:continuum}Continuum: SOC proximity Hamiltonian in magnetic field}
When graphene is brought into contact with other surfaces, internal symmetries of graphene, such as the sixfold rotational symmetry or the horizontal mirror symmetries are broken. This is reflected in the low-energy electronic structure, where the spin degeneracy of the Dirac bands is lifted and a gap can be introduced due to chiral symmetry breaking. On a microscopic level, symmetry reduction manifests in terms of extra electron hopping processes, which are forbidden in pristine graphene.

We follow the conventions introduced in Ref.~\onlinecite{Kochan2017}, defining a $C_{3v}$-symmetric, linearized effective Hamiltonian
\begin{eqnarray}
	\mathcal{H}^\kappa &=& \mathcal{H}^\kappa_{k} + \mathcal{H}_{\Delta} + \mathcal{H}^\kappa_{R} + \mathcal{H}^\kappa_{I} \label{eq:proximity_hamiltonian}\\
	\mathcal{H}^\kappa_{k} &=& \hbar v_{F} (\kappa k_x \sigma_x - k_y \sigma_y) \otimes s_0,\label{eq:graphene}\\
	\mathcal{H}_{\Delta} &=& \Delta\,\sigma_z \otimes s_0,\label{eq:staggered}\\
	\mathcal{H}^\kappa_{R} &=& \LR(-\kappa\sigma_x \otimes s_y-\sigma_y \otimes s_x), \label{eq:rashba}\\
	\mathcal{H}^\kappa_{I} &=& \frac{1}{2} \kappa \left [\LIA  (\sigma_z+\sigma_0) + \LIB(\sigma_z-\sigma_0)\right ] \otimes s_z\label{eq:intrinsic}.
\end{eqnarray}
The spin-orbit coupling proximity Hamiltonian~\eqref{eq:proximity_hamiltonian} is given for a specific valley $\kappa=\pm1$ (representing the $\K$/$\Kp$~points) and separates into the bare graphene part, Eq.~\eqref{eq:graphene}, the staggered potential part, Eq.~\eqref{eq:staggered}, the Rashba spin-orbit coupling, Eq.~\eqref{eq:rashba}, and the sublattice-resolved intrinsic spin-orbit coupling part, Eq.~\eqref{eq:intrinsic}. The bare graphene part involves the Fermi velocity $v_F=\sqrt{3}at/2\hbar$, which depends on the nearest-neighbor hopping strength $t$ and lattice constant $a$. For details of the underlying microscopic hopping processes see Refs.~\onlinecite{Gmitra2016,Kochan2017,Frank2018}. The symbols $\sigma$ and $s$ stand for unit and Pauli matrices describing sublattice and real spin degrees of freedom, respectively.

\subsection{SOC proximity Hamiltonian in orbital magnetic field}
To study the orbital effects of a transverse magnetic field on the proximitized graphene system, we apply the continuum approximation and minimal coupling substitution
\begin{equation}
	k_x \rightarrow (p_x +eA_x)/\hbar \rightarrow k_x - \epsilon\frac{e\B y}{\hbar}.
\end{equation}
The Landau gauge with vector potential $\vec{A}(\vec{r}) = (-\epsilon \B y, 0, 0)$ is applied to produce a magnetic field perpendicular to graphene. In our notation $\B$ is always the modulus of the $\B$-field and $\epsilon=\pm$ indicates its sign. Implicitly, we assume a plane wave ansatz in the $x$ direction. Following the approach of Ref.~\onlinecite{DeMartino2011}, we define bosonic Landau ladder operators $a, a^\dagger$ that lead to a replacement rule of the original matrix elements of
\begin{align}
	\hbar v_F\left(\left[\epsilon k_x+\frac{e\B y}{\hbar}\right]+ik_y\right) & \rightarrow \hbar \omega_\B a,\label{eq:lowering}\\
	\hbar v_F\left(\left[\epsilon k_x+\frac{e\B y}{\hbar}\right]-ik_y\right) & \rightarrow \hbar \omega_\B a^\dagger.\label{eq:raising}
\end{align}
We define the cyclotron frequency $\omega_\B = \sqrt{2}v_F/l_\B$ with the magnetic length as $l_\B =\sqrt{\hbar/e\B}$. The operators are the standard textbook ladder operators with $a\ket{n}=\sqrt{n}\ket{n-1}$, $a^\dagger\ket{n}=\sqrt{n+1}\ket{n+1}$, acting on harmonic oscillator eigenfunctions $\ket{n},$ with $n\in\mathbb{N}^0$. 

Substituting Eqs.~\ref{eq:lowering} and \ref{eq:raising} into Eq.~\eqref{eq:proximity_hamiltonian} leads to the following positive $\B$-field and $\K$-valley Hamiltonian
	\begin{align}
		\mathcal{H}&_{\epsilon=+\B}^{\kappa=+} = \label{eq:ladder_hamiltonian}\\
		& \begin{pmatrix}
		\Delta + \LIA & 0 & -\hbar\omega_\B a^\dagger & 2i\LR \\
		0 & \Delta - \LIA & 0 & -\hbar\omega_\B a^\dagger \\
		-\hbar\omega_\B a & 0 & -\Delta -\LIB & 0 \\
		-2i\LR & -\hbar\omega_\B a & 0 & -\Delta +\LIB
	\end{pmatrix}\nonumber,
	\end{align}
given in the basis ($A\uparrow$, $A\downarrow$, $B\uparrow$, $B\downarrow$).
To solve the Schr\"{o}dinger equation, we employ the spinor ansatz
	\begin{equation}
		\Psi^{\kappa=+}_{\epsilon=+\B,n,m} = \begin{pmatrix}
		c^{A\uparrow}_{n,m} \ket{n}\\
		c^{A\downarrow}_{n,m} \ket{n+1}\\
		c^{B\uparrow}_{n,m} \ket{n-1}\\
		c^{B\downarrow}_{n,m} \ket{n}
	\end{pmatrix}.
	\end{equation}
This ansatz represents a combination of different harmonic oscillator eigenfunctions for different spinor components, which are consistent with the matrix of Eq.~\eqref{eq:ladder_hamiltonian} acting on it. The quantum number $m$ labels the subset of solutions for orbital index $n$, with linear combination coefficients $c_{n,m}$. After acting on the ansatz, we obtain the matrix
	\begin{align}
		\mathcal{H}&^{\kappa=+}_{\epsilon = +\B, n} = \label{eq:landau_hamiltonian_K}\\
		& \begin{pmatrix}
		\Delta + \LIA & 0  & -\hbar\omega_\B \sqrt{n}  & 2i\LR  \\
		0 & \Delta - \LIA & 0 & -\hbar\omega_\B \sqrt{n+1} \\
		-\hbar\omega_\B \sqrt{n} & 0 & -\Delta -\LIB & 0 \\
		-2i\LR & -\hbar\omega_\B \sqrt{n+1} & 0 & -\Delta +\LIB
		\end{pmatrix}.\nonumber
	\end{align}
A more general expression for the other valley $\Kp$~and opposite magnetic fields can be found in the appendix.

\subsection{Low-energy Landau levels}
Hamiltonian of Eq.~\eqref{eq:landau_hamiltonian_K} for a general $n\geq1$ can be solved by numerical diagonalization. This section deals with the discussion of low orbital indices $n$ and their associated eigenstates. Harmonic oscillator eigenfunctions are only defined for quantum numbers $n\geq 0$. Here, in principle also negative orbital indices are allowed, as long as they do not produce trivial solutions, e.g., for $n=-1$, we obtain 
\begin{equation}
	\Psi^{\kappa=+}_{\epsilon = +\B, n=-1} = \begin{pmatrix}
	0\\
	\ket{0}\\
	0\\
	0\\
	\end{pmatrix},\label{eq:constant_solution}
\end{equation}
effectively projecting Hamiltonian of Eq.~\eqref{eq:landau_hamiltonian_K} to the $A\downarrow$ subspace. This leads to the Hamiltonian and eigenvalue of 
\begin{equation}
	\mathcal{H}^{\kappa=+}_{\epsilon = +\B, n=-1} = \Delta - \LIA, \label{eq:energy_ll_singlet}
\end{equation}
for a state that is completely localized in the $A$ sublattice and which is independent of magnetic field.

For the next higher orbital index $n=0$ at $\K$, one has the ansatz
\begin{equation}
\Psi^{\kappa=+}_{\epsilon = +\B, n=0, m} = \begin{pmatrix}
	c^{A\uparrow}_{0,m}\ket{0}\\
	c^{A\downarrow}_{0,m}\ket{1}\\
	0\\
	c^{B\downarrow}_{0,m}\ket{0}
\end{pmatrix},
\end{equation}
leading to an effective $3\times 3$ Hamiltonian
\begin{equation}
	\mathcal{H}^{\kappa=+}_{\epsilon = +\B, n=0} =
	\begin{pmatrix} \Delta + \LIA & 0 & 2i\LR \\
	0 & \Delta - \LIA   & -\hbar\omega_\B \\
	-2i\LR & -\hbar\omega_\B & -\Delta +\LIB
	\end{pmatrix}.
	\label{eq:landau_hamiltonian_n0}
\end{equation}
Analytical diagonalization of Hamiltonian of Eq.~\eqref{eq:landau_hamiltonian_n0} is possible, but solutions are too lengthy to be given here.

In the following we also add a Zeeman term
\begin{equation}
	\mathcal{H}_{Z} = \epsilon g_s \mu_B  \B \sigma_0 \otimes s_z, \label{eq:zeeman}
\end{equation}
with $\mu_B= e\hbar/2m_e$, the Bohr magneton and electron spin g-factor $g_s$.

Pure graphene under the action of an orbital magnetic field has four degenerate zero-energy states, two for each spin and two for each valley. If the proximity parameters of graphene tend to zero, these zero energy modes should be reproduced. So far we found one of these states with Eq.~\eqref{eq:constant_solution}. Another one turns out to be included in the $n=0$ space, by analyzing the dependence of eigenenergies of Hamiltonian of Eq.~\eqref{eq:landau_hamiltonian_n0} with respect to large magnetic fields to order $\mathcal{O}(1/\B)$. Including also Zeeman energy and carrying out the analysis for the other valley, this leads to the four low-energy states $E^{\kappa \sigma s}_{\epsilon}$
\begin{align}
	E^{+ A \uparrow}_{+} &= \Delta + \LIA + g_s \mu_B \B + \mathcal{O}\left(\frac{1}{\B}\right),\label{eq:ll1}\\
	E^{+ A \downarrow}_{+} &= \Delta - \LIA - g_s \mu_B \B, \label{eq:ll2}\\
	E^{- B \uparrow}_{+} &= -\Delta + \LIB + g_s \mu_B \B + \mathcal{O}\left(\frac{1}{\B}\right),\label{eq:ll3}\\
	E^{- B \downarrow}_{+} &= -\Delta - \LIB - g_s \mu_B \B.\label{eq:ll4}
\end{align}

Interestingly, the $\B$-field selects states in the $A$ sublattice in valley $\kappa=+$ and from $B$ sublattice in valley $\kappa=-$, as indicated by the superscripts. Even without the Zeeman term, an orbital magnetic field is able to break the Kramers degeneracy of levels by coupling to SOC hopping processes, which is not possible, for example, in pure graphene (without SOC). Equations~\eqref{eq:ll1}--\eqref{eq:ll4} could be used in experiment to extract $\Delta$, $\LIA$, and $\LIB$. 

For negative magnetic fields ($\epsilon=-$), creation and annihilation operators are interchanged ($a\leftrightarrow-a^\dagger$) in Eq.~\eqref{eq:ladder_hamiltonian}, leading to a different $n$-dependent Hamiltonian $\mathcal{H}^{\kappa}_{\epsilon=-\B,n}$ with accordingly adapted wave function ansatz (see also appendix).

We find the symmetry relation
\begin{equation}
	\mathcal{H}_{-\epsilon \B,n}^{-\kappa} = (\sigma_0 \otimes s_x)^\dagger~\mathcal{H}^{\kappa}_{\epsilon \B,n}~(\sigma_0 \otimes s_x),
\end{equation}	
which leads to a connection between levels at opposite magnetic fields
\begin{equation}
	E_{-\epsilon \B}^{-\kappa \sigma -s} = E_{\epsilon \B}^{\kappa \sigma s},
\end{equation}
i.e. for each positive magnetic field $\B$, there is a corresponding level at $-\B$ with the same energy, but with flipped valley and spin quantum numbers, in line with the operation of time reversal. 

\begin{figure*}
	\centering
	\includegraphics[width=0.9\textwidth]{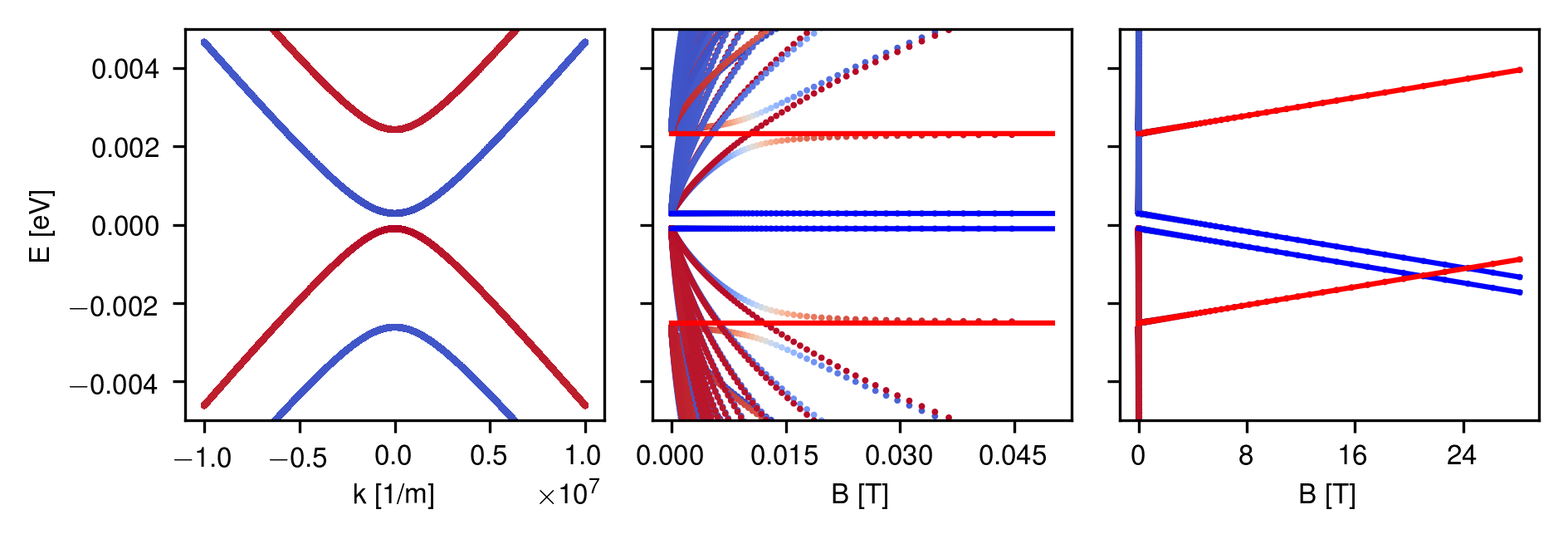}
	 \caption{\label{fig:landau_ws2}Bulk and Landau level spectrum for graphene/WS$_2$ for small and large values of the magnetic field $\B$. The bulk spectrum is from the $\K$ point. Straight lines are plots of the low-energy Landau level formulas. Color code corresponds to $s_z$ expectation values (red is spin up, blue is spin down). Parameters: $t=-2.657$~eV, $\Delta=1.31$~meV, $\LR=0.36$~meV, $\LIA=1.02$~meV, $\LIB=-1.21$~meV.}
	\vspace*{\floatsep}
	\includegraphics[width=0.9\textwidth]{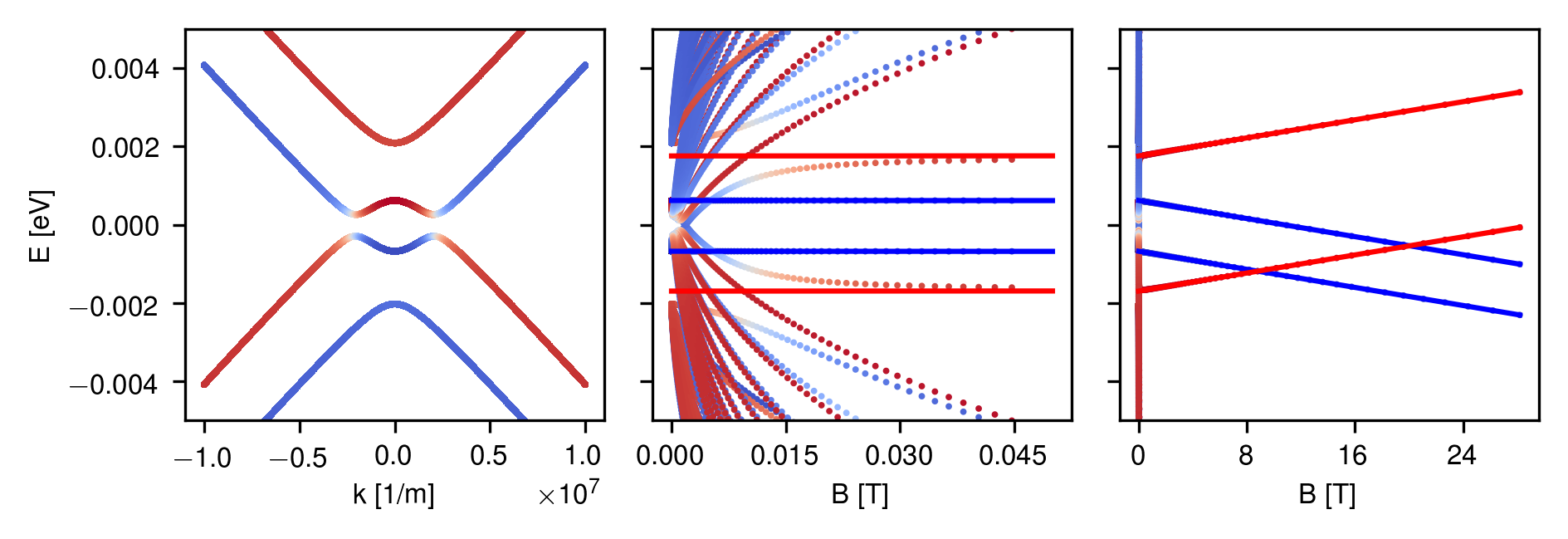}
 \caption{\label{fig:landau_wse2}Bulk and Landau level spectrum for graphene/WSe$_2$ for small and large values of the magnetic field $\B$. The bulk spectrum is from the $\K$ point. Straight lines are plots of the low-energy Landau level formulas. Color code corresponds to $s_z$ expectation values (red is spin up, blue is spin down). Parameters: $t=-2.507$~eV, $\Delta=0.54$~meV, $\LR=0.56$~meV, $\LIA=1.22$~meV, $\LIB=-1.16$~meV.}
\end{figure*}

\subsection{Realistic parameter values for graphene on TMDs}

A general motivation for this study is to develop techniques to extract parameters in conjunction with local magneto-spectroscopy. In experiment, spin-orbit coupling and orbital model parameters may vary spatially due to formation of moir\'{e} patterns originating from a twist angle between graphene and the substrate or due to lattice mismatch. In this section we show which types of local fan charts can be expected and whether fingerprints of specific parameter regimes can be recognized therein.

To this end, we take \emph{ab-initio} extracted parameters for commensurate system combinations~\cite{Gmitra2016} and compare spin-resolved eigenvalues of Hamiltonian of Eq.~\eqref{eq:landau_hamiltonian_K} at $\kappa=\pm $ in Figs.~\ref{fig:landau_ws2} and \ref{fig:landau_wse2} versus increasing magnetic field [including the Zeeman term of Eq.~\eqref{eq:zeeman}] with the bulk band structure.

In the case of graphene on WS$_2$, where $\LIA\approx-\LIB$, $\LIA < \Delta$, $\LR<\LIA$, i.e. the non-inverted regime, we see that for small magnetic field values, the Landau levels approach exactly the eigenspectrum at the $\K$ point of the bulk spectrum. Further, for small magnetic field values, Zeeman energy is not important and we find two constant eigenvalues of spin-down polarization, as indicated by formulas~\eqref{eq:ll2} and~\eqref{eq:ll4}. The other two eigenvalues \eqref{eq:ll1} and \eqref{eq:ll3} are represented already at very small magnetic fields of about 0.03~T. Zeeman energy and the presence of the linear-in-$\B$ low-energy Landau levels become evident for large magnetic fields. It is surprising that spin degeneracy is almost immediately broken by the magnetic field, where it selects only the innermost Landau levels to have spin-down polarization of different valleys. This is inverted by application of negative magnetic fields (not shown). The second higher Landau level shows an anticrossing, which mixes spin-up and down species and relates to the Rashba spin-orbit interaction. For high magnetic fields (above 20~T), we find the low-energy Landau levels to cross, which could be used to extract model parameters with the help of Eqs.~\eqref{eq:ll1}--\eqref{eq:ll4}.

In the inverted case of graphene on WSe$_2$, for which $\LIA\approx-\LIB$, $\LIA > \Delta$, $\LR<\LIA$, i.e., intrinsic spin-orbit coupling dominates all energy scales, the Landau level structure is very similar to the previous case except for very small magnetic fields in the mT regime. We observe a very strong response to the magnetic field where two levels are crossing at a certain critical field and the system is \textit{no longer} gapped for all magnetic fields. This will be discussed in more detail in the next section.

\begin{figure*}
	\centering
	\includegraphics[width=0.9\textwidth]{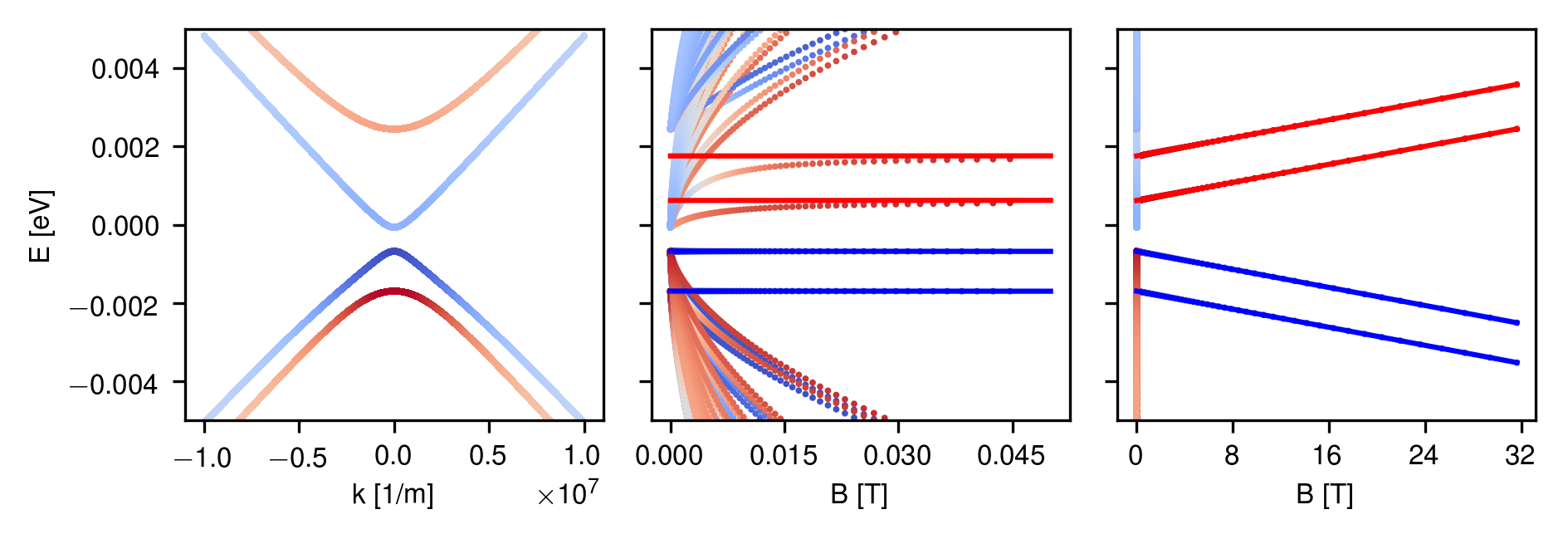}
	 \caption{\label{fig:landau_wse2_km}Bulk and Landau level spectrum for graphene/Kane-Mele-WSe$_2$ with $\LIA=\LIB$ for small and large values of the magnetic field $\B$. The bulk spectrum is from the $\K$ point. Straight lines are plots of the low-energy Landau level formulas. Color code corresponds to $s_z$ expectation values (red is spin up, blue is spin down). Parameters: $t=-2.507$~eV, $\Delta=0.54$~meV, $\LR=0.56$~meV, $\LIA=1.22$~meV, $\LIB=1.16$~meV.}
	\vspace*{\floatsep}
	\includegraphics[width=0.9\textwidth]{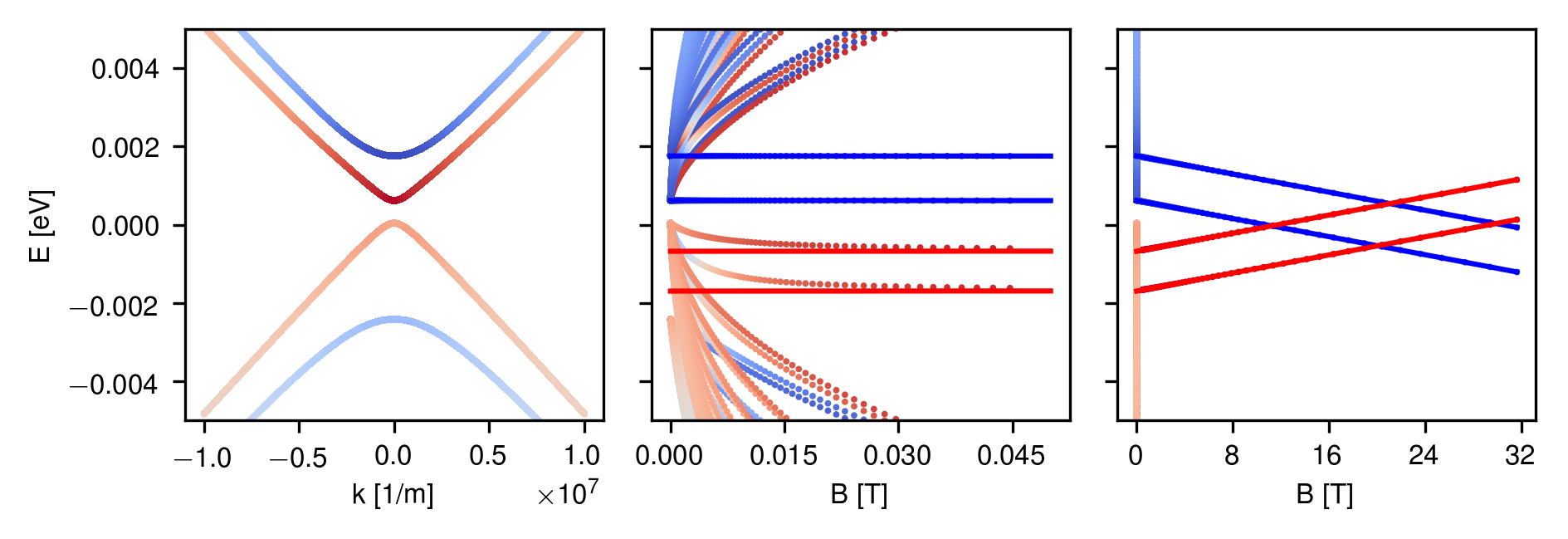}
	 \caption{\label{fig:landau_wse2_nkm}Bulk and Landau level spectrum for graphene/Kane-Mele-WSe$_2$ with $\LIA=\LIB$ for small and large values of the magnetic field $\B$. The bulk spectrum is from the $\K$ point. Straight lines are plots of the low-energy Landau level formulas. Color code corresponds to $s_z$ expectation values (red is spin up, blue is spin down). Parameters: $t=-2.507$~eV, $\Delta=0.54$~meV, $\LR=0.56$~meV, $\LIA=-1.22$~meV, $\LIB=-1.16$~meV.}
\end{figure*}

Moreover, we apply a set of parameters, which are the same as for WSe$_2$, but $\LIB \rightarrow -\LIB$, resulting in about equal sized intrinsic SOC parameters. The spectra for this situation are shown in Fig.~\ref{fig:landau_wse2_km}. This large uniform intrinsic SOC results in a quantum spin Hall topological insulator~\cite{Kane2005}. Even for large magnetic fields, we see no crossing of energy levels at zero magnetic fields in Fig.~\ref{fig:landau_wse2_km}. Conversely, having a negative sign for the intrinsic spin-orbit couplings results in two crossings at about zero energy, see Fig.~\ref{fig:landau_wse2_nkm}. In literature commonly only the case of negative intrinsic spin-orbit couplings (in our convention) is studied~\cite{DeMartino2011}. The number of crossings and the symmetry with respect to energy could be used to qualitatively determine the relative and absolute signs of the intrinsic SOC parameters and therefore provide information about topologically trivial and nontrivial regimes. Electron-hole symmetry in the fan chart is a fingerprint of uniform, Kane-Mele SOC, absence of electron-hole symmetry points to staggered, valley Zeeman type of intrinsic SOC. 

\subsection{Discussion of gap closing\label{sec:gap_closing}}
Gap inversion in graphene on WSe$_2$ leads to very curious effects even without magnetic fields. Pseudohelical edge states can appear which are as robust as topological states from a topological insulator~\cite{Frank2018}. The inverted regime behaves peculiar with magnetic field as well; we observe a gap closing in Fig.~\ref{fig:landau_wse2} as compared to the preserved gap in WS$_2$. This gap closing is not related to the Zeeman energy but is a purely orbital effect.

\begin{figure}[h]
    \centering
    \includegraphics{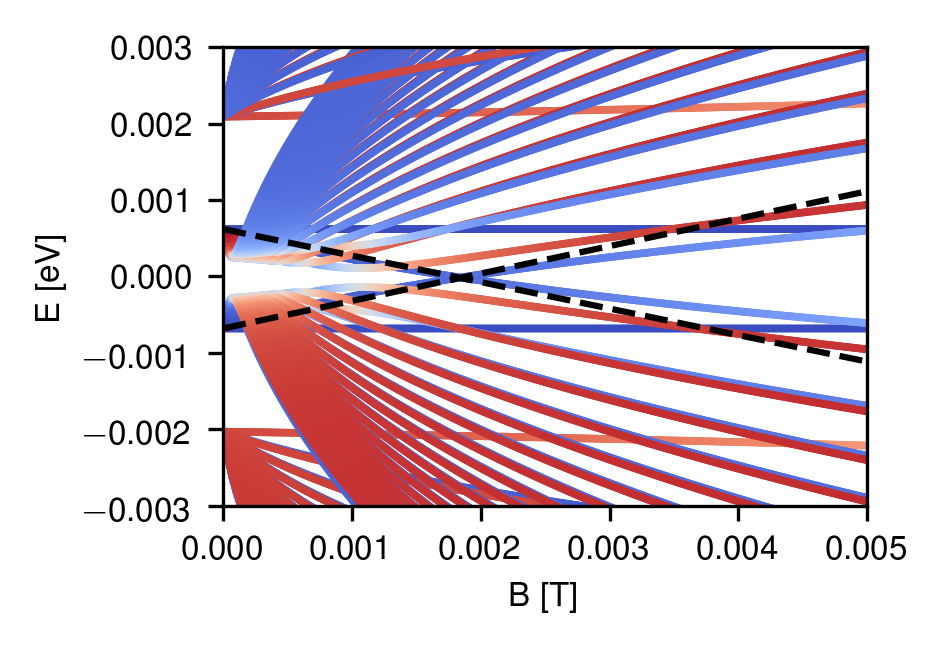}
    \caption{Zoom into the low $\B$-field and low-energy Landau level states of graphene on WSe$_2$ of Fig.~\ref{fig:landau_wse2}. Dashed lines are eigenvalues for downfolded Hamiltonians onto subspaces of the $n=0$ Landau Hamiltonian.}
    \label{fig:landau_wse2_zoom}
\end{figure}

In Fig.~\ref{fig:landau_wse2_zoom} we zoom into the low-$\B$ physics of Fig.~\ref{fig:landau_wse2}. From our analysis we can see that only exactly one state from each valley is able to cross without interaction close to zero energy. One of the states stems from the $3\times 3$ Hamiltonian of Eq.~\eqref{eq:landau_hamiltonian_n0}, the other from the other valley counterpart. The states in question are the $A\downarrow$ from $\kappa=+$ and $B\downarrow$ from $\kappa=-$. By L\"{o}wdin-downfolding the $3\times3$ Hamiltonians for $n=0$ at energy $E=0$ to these subspaces, an approximation to the eigenvalue can be obtained in both cases, leading to expressions
\begin{align}
	\mathcal{H}^{\kappa=+}_\mathrm{eff} &= \Delta -\LIA + \frac{(\Delta+\LIA) \lambda_v^2 \B}{(\Delta+\LIA)(\Delta-\LIB) + 4\LR^2},\label{eq:crossing_eigenvalue_k} \\
	\mathcal{H}^{\kappa=-}_\mathrm{eff} &= -\Delta -\LIB - \frac{(\Delta-\LIB) \lambda_v^2 \B}{(\Delta+\LIA)(\Delta-\LIB) + 4\LR^2},\label{eq:crossing_eigenvalue_kp}
\end{align}
where $\lambda_v= \hbar\omega_\B/\sqrt{\B}=\sqrt{2e\hbar} v_F$.

These expressions only contain linear terms in $\B$. This is due to the downfolding procedure, which neglects higher order terms. Equations~\eqref{eq:crossing_eigenvalue_k} and \eqref{eq:crossing_eigenvalue_kp} are plotted in Fig.~\ref{fig:landau_wse2_zoom}. Starting at the exact eigenvalues of the bulk spectrum at the $\K$/$\Kp$ points, they cross at a certain magnetic field, at about the same point where the numerical data intersects. We tested this to hold for different combinations of parameters in the inverted regime. Thus, we can combine Eqs.~\eqref{eq:crossing_eigenvalue_k} and \eqref{eq:crossing_eigenvalue_kp} to estimate the critical magnetic field
\begin{equation}
	\B_\mathrm{crit} = \frac{(\LIA - \LIB - 2\Delta)((\Delta+\LIA)(\Delta-\LIB) + 4\LR^2)}{\lambda_v^2(2\Delta + \LIA - \LIB)},\label{eq:crit_field}
\end{equation}
for which we expect the inner states to cross. Inserting model values for WSe$_2$, we obtain a critical field of 1.8~mT.

While being very small, the critical field scales quadratically with the intrinsic SOC. For tenfold increased intrinsic SOC, predicted by some experiments~\cite{Garcia2018}, one expects critical field values on the order of 100~mT. Formula~\eqref{eq:crit_field} depends on the Rashba parameter, which offers a way to extract the parameter by first determining the other parameters with the low-energy Landau level expressions of Eqs.~\eqref{eq:ll1}--\eqref{eq:ll4} and then plugging in the critical magnetic field.

Further, Eqs.~\eqref{eq:crossing_eigenvalue_k} and \eqref{eq:crossing_eigenvalue_kp} can be interpreted in terms of a Zeeman-like (linear-in-$\B$) response. Magnetic moments at $\K$~and $\Kp$~can be identified as
\begin{equation}
	m_{\K/\Kp} = \pm \frac{\lambda_v^2(\Delta\pm\lambda_I^{A/B})}{(\Delta+\LIA)(\Delta-\LIB) + 4\LR^2},\label{eq:magnetic_moments_AB}
\end{equation}
which for staggered $\LIA = -\LIB = \LI$ simplifies to
\begin{equation}
	m_{\K/\Kp} = \pm \frac{\lambda_v^2(\Delta+\LI)}{(\Delta+\LI)^2 + 4\LR^2}.\label{eq:magnetic_moments}
\end{equation}
Inserting WSe$_2$ parameters, we obtain $m\approx\pm 0.4~\mathrm{eV/T} = 7256~\mu_B$. Interestingly, for $\LI=\LR=0$, this formula reduces to four times the magnetic moment found by Xiao et al.~\cite{Xiao2007} and thus can be brought into connection with the self-rotating magnetic moment discussed in Sec.~\ref{sec:magnetic_moments}. The magnetic moment of Eq.~\eqref{eq:magnetic_moments} is only an approximation as it follows from a linearized Hamiltonian, but has the correct qualitative behavior, namely it scales quadratically with the velocity and is inversely proportional to the energy gap; compare to formula \eqref{eq:self-rotating} in Sec.~\ref{sec:magnetic_moments}.

\section{\label{sec:magnetic_moments}Magnetic moments}
The response to orbital magnetic fields in this system is very strong, particularly in the inverted-band case. Here, we want to pick up the idea of magnetic moments appearing in Eq.~\eqref{eq:magnetic_moments_AB}. It is known that inversion symmetry breaking in graphene, as experienced in proximitized graphene, can have consequences for the semiclassical motion of electrons~\cite{Xiao2007}. Specifically, it manifests in a valley Hall effect, which generates a transversal valley current, i.e., local imbalance of valley population, a consequence of valley-dependent Berry curvature. This effect is very similar to the intrinsic spin Hall effect which originates from a spin-dependent Berry curvature. Berry curvature around graphene valleys can have very peculiar shapes in these systems and determines their topological properties~\cite{Frank2018}.

Inversion symmetry allows also for finite magnetic moments in reciprocal space $\vec{m}_{n\vec{k}}$, dependent on the band and k vector of the Bloch state. This magnetic moment contribution plays an important role in the formation of the macroscopic orbital magnetization of solids~\cite{Thonhauser2005} and should be detectable in experiment~\cite{Xiao2007}. It also has implications for the semiclassical single-particle band energy of electrons, which gets modified by an additional Zeeman-like term $\varepsilon^\B_{n\vec{k}} = \varepsilon_{n\vec{k}} - m_{n\vec{k}} \B$~\cite{Xiao2010}. 

The often called self-rotating magnetic moment, can be expressed in terms of perturbation theory~\cite{Chang1996}
\begin{align}
	&m_{n \vec{k}}^z = -i \frac{e}{2\hbar}\bra{\partial_{k_x} u_{n\vec{k}}} \mathcal{H}_\vec{k} - E_{n\vec{k}} \ket{\partial_{k_y} u_{n\vec{k}}} - x\leftrightarrow y \nonumber \\
	& = i \frac{e}{2\hbar}\sum_{l\neq n} \frac{\bra{u_{n\vec{k}}} \frac{\partial \mathcal{H}}{\partial k_x} \ket{u_{l\vec{k}}}\bra{u_{l\vec{k}}} \frac{\partial \mathcal{H}}{\partial k_y} \ket{u_{n\vec{k}}}}{E_{n\vec{k}}-E_{l\vec{k}}} - x\leftrightarrow y.\label{eq:self-rotating}
\end{align}

Without time-reversal symmetry breaking, $m_{n \vec{k}}^z$ transforms like angular momentum or Berry curvature and is odd in $\vec{k}$. This is in nice agreement with formulas~\eqref{eq:crossing_eigenvalue_k} \eqref{eq:crossing_eigenvalue_kp}, and Eq.~\eqref{eq:magnetic_moments}, reflecting the different signs in the different valleys.

\begin{figure}
	\subfloat[WS$_2$]{\includegraphics[width=0.98\columnwidth]{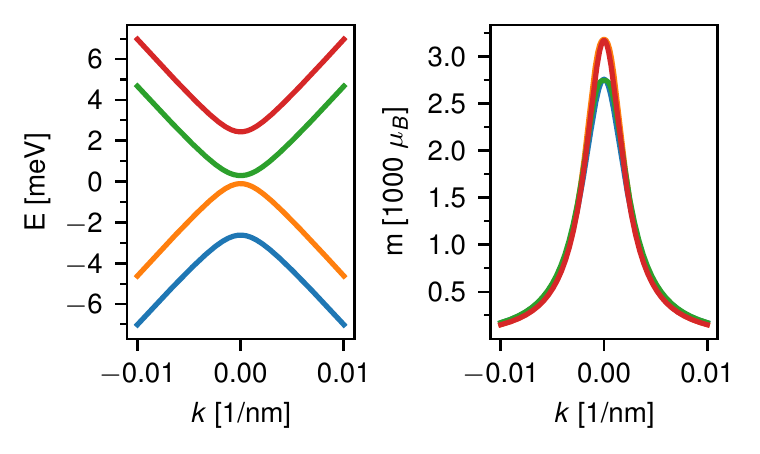}\label{fig:moment_ws2}}\\
	\subfloat[WSe$_2$]{\includegraphics[width=0.98\columnwidth]{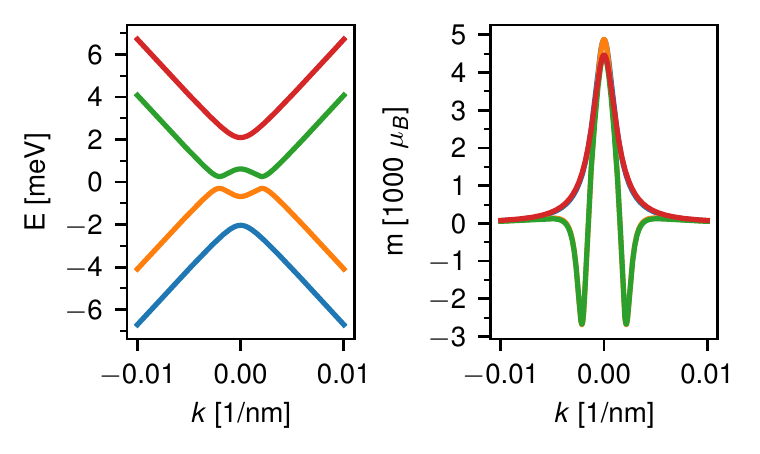}\label{fig:moment_wse2}}\\
	\caption{Low-energy band structures (left) around the $\K$ point of proximitized graphene (on WS$_2$ and WSe$_2$) with their associated magnetic moments (right). Colors of the bands correspond to band indices which are in correspondence in both figures.}
	\label{fig:magnetic_moments}
\end{figure}

We evaluate Eq.~\eqref{eq:self-rotating} for DFT-extracted model parameters~\cite{Gmitra2016} in Fig.~\ref{fig:magnetic_moments} next to their bulk band structures for the non-inverted and inverted systems graphene on WS$_2$ and WSe$_2$, respectively. For the non-inverted case, see Fig.~\subref*{fig:moment_ws2}, we find same signs of magnetic moments for all the bands in one valley. The magnitudes of the magnetic moments are exceptionally large compared to the Bohr magneton~$\mu_B$, enhanced by a factor of up to 3000. The large size of the magnetic moment can be made plausible by estimating $m/\mu_B \approx m_e v_F^2 / \Delta E \approx 10^4$, using $v_F\approx10^6$~m/s and inserting a typical gap energy scale of $\Delta E = 1$~meV. The asymmetry between the four bands stems from the slight deviations in magnitude between the intrinsic spin-orbit coupling parameters.

The inverted case, Fig.~\subref*{fig:moment_wse2} is different. Directly at the $\K$ point, we again see the behavior of only positive magnetic moments, which could have been guessed from the previous result, the bands are just inverted with respect to each other. Interactions among the bands are slightly altered compared to WS$_2$, changing the overall magnitude. We see magnetic moments of almost 5000~$\mu_B$, which is in the same order of magnitude as the results from the approximate formula~\eqref{eq:magnetic_moments}. However, directly at the anticrossings, which are generated by Rashba spin-orbit coupling, negative magnetic moments are formed. These negative magnetic moments could be a fingerprint of the inverted gap regime in experiments.

\section{\label{sec:conclusion}Conclusion}
In this work we show that Landau level spectroscopy in proximitized graphene can be an alternative way to directly measure the local magnitude of orbital and spin-orbit coupling parameters. From the Landau level fan diagrams it is possible to distinguish the nature (relative and absolute signs) of intrinsic spin-orbit couplings by counting the crossings of low-energy Landau levels and by analysis of their electron-hole symmetry. Further, the inverted band structure in graphene on WSe$_2$ opens up a new regime, where the bulk gap is not preserved for all magnetic field values. This could have potential consequences for the edge state physics in these systems, which remains to be studied. Interestingly, the response of the crossing Landau levels to magnetic fields is in the same order of magnitude as the self-rotating magnetic moments generated in the valleys of proximitized graphene. The magnetic moments can be giant, about several 1000~$\mu_B$ compared to the electron spin g-factor. The direct relation between the response of Landau levels and the magnetic moments is subject to future work.

\begin{acknowledgments}
This work was funded by the Deutsche Forschungsgemeinschaft (DFG, German Research Foundation) SFB 1277 (Project-ID 314695032). The authors gratefully acknowledge the Gauss Centre for Supercomputing e.V. (www.gauss-centre.eu) for funding this project by providing computing time on the GCS Supercomputer SuperMUC at Leibniz Supercomputing Centre (LRZ, www.lrz.de).
\end{acknowledgments}

\appendix
\section{General magnetic field Hamiltonian for field directions and valleys}

For a general valley $\kappa=\pm 1$ and sign of magnetic field $\epsilon=\pm 1$, the effective Hamiltonian matrix for orbital number $n$ is given by
\begin{align}
	& \mathcal{H}_{\epsilon \B, n}^\kappa = \mathcal{H}_\Delta + \mathcal{H}_R^\kappa + \mathcal{H}_I^\kappa + \epsilon g_s \mu_B  \B~ \sigma_0 \otimes s_z \label{eq:general_hamiltonian}\\
	& - \epsilon\kappa\hbar\omega_\B ~\sigma_x \otimes \begin{pmatrix}
		\sqrt{n+(1-\epsilon)/2} & 0 \nonumber\\
                0 & \sqrt{n+(1+\epsilon)/2}
	\end{pmatrix},
\end{align}
following the same derivation as in Sec.~\ref{sec:continuum}. The Hamiltonian of Eq.~\eqref{eq:general_hamiltonian} is only well defined in combination with the wave function ansatz it acts uppon, which are listed in Tab.~\ref{tab:hamiltonian_parameters}.

\begin{table}
	\caption{\label{tab:hamiltonian_parameters}Wave function spinor component ansatzes for the general Hamiltonian of Eq.~\eqref{eq:general_hamiltonian} for valley $\kappa$ and sign of magnetic field $\epsilon$ in the basis $(A\uparrow,A\downarrow,B\uparrow,B\downarrow)$.}
	\begin{ruledtabular}
		\begin{tabular}{llcc}
			$\epsilon$ & $\kappa$ & spinor components & constant solution\\
			\hline
			$+$ & $+$ & $\ket{n}, \ket{n+1}, \ket{n-1}, \ket{n}$ & $A\downarrow$\\
			$+$ & $-$ & $\ket{n-1}, \ket{n}, \ket{n}, \ket{n+1}$ & $B\downarrow$\\
			$-$ & $+$ & $\ket{n}, \ket{n-1}, \ket{n+1}, \ket{n}$ & $B\uparrow$\\
			$-$ & $-$ & $\ket{n+1}, \ket{n}, \ket{n}, \ket{n-1}$ & $A\uparrow$\\
		\end{tabular}
	\end{ruledtabular}
\end{table}

\bibliography{paper}

\end{document}